# A Dual Role for Sex?

Pavel Gorodetsky[1] and Emmanuel Tannenbaum[2]*

*[1]Department of Physics, Ben-Gurion University of the Negev, Be'er-Sheva, Israel; [2]Department of Chemistry, Ben-Gurion University of the Negev, Be'er-Sheva, Israel.*

[1]Electronic address: pavelgor@bgu.ac.il; [2]Electronic address: emanuelt@bgu.ac.il

**The two classic theories for the existence of sexual replication are that sex purges deleterious mutations from a population[1-7], and that sex allows a population to adapt more rapidly to changing environments[8-11]. These two theories have often been presented as opposing explanations for the existence of sex. Here, we develop and analyze evolutionary models based on the asexual and sexual replication pathways in *Saccharomyces cerevisiae* (Baker's yeast)[12-15], and show that sexual replication can *both* purge deleterious mutations in a static environment, as well as lead to faster adaptation in a dynamic environment. This implies that sex can serve a dual role, which is in sharp contrast to previous theories.**

The evolution and maintenance of sexual replication as the preferred, and in some cases only, mode of replication in complex organisms is one of the central questions of evolutionary biology. Numerous studies have been conducted on the subject of sexual replication, some of which have argued that sex is advantageous because it helps to remove deleterious mutations[1-7], and some of which have argued that sex allows for faster adaptation in dynamic environments[8-11]. These two classes of theories have often been presented as opposing explanations for the existence of sex. Here, we develop mathematical models based on the asexual and sexual replication pathways of *Saccharomyces cerevisiae*, or Baker's yeast[12-15], in order to test the plausibility of both hypotheses for the existence of sex for the case of unicellular organisms. We find, in contrast to previous explanation for the existence of sex, that sex can serve a dual role: That is, sex may both act as a mechanism for purging deleterious mutations in a static environment, as well as allow a population to adapt more quickly in a time-varying environment.

In order to analyze asexual and sexual replication in unicellular organisms, we begin by assuming a unicellular population of replicating organisms, where each organism has a two-chromosome genome, and each chromosome is taken to be of length $L$, where each base is chosen from an alphabet of size $S$ (for simplicity we are assuming single-stranded genomes). A chromosome is taken to be *viable* if its base sequence is identical to some "master" sequence $\sigma_0$. Otherwise, the chromosome is said to be *unviable*. The first-order growth rate constant, or *fitness*, of an organism is then determined by the number of viable and unviable chromosomes in its genome. $\kappa_{vv}$ denotes the fitness of an organism with two viable chromosomes, $\kappa_{vu}$ denotes the fitness of an organism with one viable and one unviable chromosome, and $\kappa_{uu}$ denotes the fitness of an organism with two unviable chromosomes. It is assumed that $\kappa_{vv} > \kappa_{vu} > \kappa_{uu}$. It should be noted that, in the language of population genetics, we are considering what is known as a *single-locus model*.

We also assume that the master sequence shifts by a single point mutation after every time $\tau$. It is assumed that the point mutation occurs in an un-mutated portion of the genome, so that the master sequences move further away in sequence space from the original master sequence[16]. A static landscape is obtained by taking the limit $\tau \to \infty$.

During the mitotic cell division for the case of asexual replication, each of the parent chromosomes replicate, and the resulting pairs segregate into the daughter cells, where each cell receives one of the chromosomes from each pair. It is assumed that replication is not error-free, so that there is a per-base error probability $\epsilon$.

For sexual replication, the diploid cell divides into two diploids via the normal mitotic pathway described above, after which each of the diploids divide into two haploids, producing a total of four haploids[12-15]. These haploids enter a haploid pool, where they fuse with other haploids to produce diploids that then repeat the life cycle. It is assumed that haploid fusion is random, and is characterized by a second-order rate constant γ. Furthermore, it is assumed that the population density ρ, defined as the total number of diploids plus half the number of haploids, divided by the total volume, remains constant.

The evolution of the asexual and sexual populations may be simulated stochastically in a chemostat-type model[17], where a fixed total population of organisms undergoes replication and mutation according to probabilities defined by the first-order growth rate constants and error probabilities defined above. The per organism replication probability per time step is $\kappa_{rs=vv,vu,uu} \Delta t$, where $\kappa_{rs}$ is the fitness of the organism, and $\Delta t$ is the simulation time step. For the sexual case, it is also necessary to consider haploid fusion, which occurs with a pairwise probability that is given by $(\frac{\gamma}{V})\Delta t$. Organisms are randomly removed at each time step in order to maintain a constant population size.

The evolutionary dynamics of the asexual and sexual populations may each be accurately described by respective systems of ordinary differential equations. For the asexual population, we define $n_{vv}$ to be the number of organisms with two viable chromosomes, $n_{vv_1}$ to be the number of organisms with one viable chromosome and one chromosome that will be viable after the next peak shift, $n_{vu}$ to be the number of organisms with one viable and one unviable chromosome, $n_{v_1v_1}$ to be the number of organisms where both chromosomes will be viable after the next peak shift, $n_{v_1u}$ to be the number of organisms where one chromosome will be viable after the next peak shift and one chromosome is unviable, and $n_{uu}$ to be the number of organisms with two unviable chromosomes (note that a "$v_1$" chromosome is also unviable, but we place it in a separate category from the other unviable chromosomes, since it will become viable after one peak shift). For the sexual population, the diploid population numbers are defined as for the asexual population.

In addition, we also define haploid population numbers $n_v, n_{v_1}, n_u$.

The probability that a given chromosome is replicated correctly is $p \equiv (1-\epsilon)^L$, and the probability that a given chromosome produces a daughter with a specific sequence that differs from the parent by a single point mutation is $q \equiv (\frac{\epsilon}{S-1})(1-\epsilon)^{L-1}$. We also assume that the genomes are sufficiently long that backmutations may be neglected, so that a "$u$" parent chromosome only produces "$u$" daughters.

With these approximations, the asexual population is governed by the following system of ordinary differential equations:

$$\frac{d n_{vv}}{dt} = [\frac{1}{2}(1+p)^2 - 1]\kappa_{vv} n_{vv} + \frac{1}{2}q(1+p)\kappa_{vu} n_{vv_1} + \frac{1}{2}q^2 \kappa_{uu} n_{v_1 v_1}$$

$$\frac{d n_{vv_1}}{dt} = q(1+p)\kappa_{vv} n_{vv} + [\frac{1}{2}((1+p)^2 + q^2) - 1]\kappa_{vu} n_{vv_1} + q(1+p)\kappa_{uu} n_{v_1 v_1}$$

$$\frac{d n_{v_1 v_1}}{dt} = \frac{1}{2}q^2 \kappa_{vv} n_{vv} + \frac{1}{2}q(1+p)\kappa_{vu} n_{vv_1} + [\frac{1}{2}(1+p)^2 - 1]\kappa_{uu} n_{v_1 v_1}$$

(1) $$\frac{d n_{vu}}{dt} = (1-p-q)(1+p)\kappa_{vv} n_{vv} + \frac{1}{2}[1-(p+q)^2]\kappa_{vu} n_{vv_1} + q(1-p-q)\kappa_{uu} n_{v_1 v_1}$$
$$+ p\kappa_{vu} n_{vu} + q\kappa_{uu} n_{v_1 u}$$

$$\frac{d n_{v_1 u}}{dt} = q(1-p-q)\kappa_{vv} n_{vv} + \frac{1}{2}[1-(p+q)^2]\kappa_{vu} n_{vv_1} + (1-p-q)(1+p)\kappa_{uu} n_{v_1 v_1}$$
$$+ q\kappa_{vu} n_{vu} + p\kappa_{uu} n_{v_1 u}$$

$$\frac{d n_{uu}}{dt} = \frac{1}{2}(1-p-q)^2 \kappa_{vv} n_{vv} + \frac{1}{2}(1-p-q)^2 \kappa_{vu} n_{vv_1} + \frac{1}{2}(1-p-q)^2 \kappa_{uu} n_{v_1 v_1}$$
$$+ (1-p-q)\kappa_{vu} n_{vu} + (1-p-q)\kappa_{uu} n_{v_1 u} + \kappa_{uu} n_{uu}$$

while the sexual population is governed by the following system of ordinary differential equations:

(2)
$$\frac{d n_{vv}}{dt} = -\kappa_{vv} n_{vv} + \frac{1}{2}\frac{\gamma\rho}{n} n_v^2$$

$$\frac{d n_{vv_1}}{dt} = -\kappa_{vu} n_{vv_1} + \frac{\gamma\rho}{n} n_v n_{v_1}$$

$$\frac{d n_{vu}}{dt} = -\kappa_{vu} n_{vu} + \frac{\gamma\rho}{n} n_v n_u$$

$$\frac{d n_{v_1 v_1}}{dt} = -\kappa_{uu} n_{v_1 v_1} + \frac{1}{2}\frac{\gamma\rho}{n} n_{v_1}^2$$

$$\frac{d n_{v_1 u}}{dt} = -\kappa_{uu} n_{v_1 u} + \frac{\gamma\rho}{n} n_{v_1} n_u$$

$$\frac{d n_{uu}}{dt} = -\kappa_{uu} n_{uu} + \frac{1}{2}\frac{\gamma\rho}{n} n_u^2$$

$$\frac{d n_v}{dt} = 2(1+p)\kappa_{vv} n_{vv} + (1+p+q)\kappa_{vu} n_{vv_1} + (1+p)\kappa_{vu} n_{vu} + 2q\kappa_{uu} n_{v_1 v_1} + q\kappa_{uu} n_{v_1 u}$$
$$- \frac{\gamma\rho}{n} n_v (n_v + n_{v_1} + n_u)$$

$$\frac{d n_{v_1}}{dt} = 2q\kappa_{vv} n_{vv} + (1+p+q)\kappa_{vu} n_{vv_1} + q\kappa_{vu} n_{vu} + 2(1+p)\kappa_{uu} n_{v_1 v_1} + (1+p)\kappa_{uu} n_{v_1 u}$$
$$- \frac{\gamma\rho}{n} n_{v_1} (n_v + n_{v_1} + n_u)$$

$$\frac{d n_u}{dt} = 2(1-p-q)\kappa_{vv} n_{vv} + 2(1-p-q)\kappa_{vu} n_{vv_1} + (3-p-q)\kappa_{vu} n_{vu} + 2(1-p-q)\kappa_{uu} n_{v_1 v_1}$$
$$+ (3-p-q)\kappa_{uu} n_{v_1 u} + 4\kappa_{uu} n_{uu} - \frac{\gamma\rho}{n} n_u (n_v + n_{v_1} + n_u)$$

After a peak shift at time $k\tau$, the population numbers are re-defined via,

(3)
$$n_{vv}(k\tau) = \lim_{t \to k\tau, t < k\tau} n_{v_1 v_1}(t)$$
$$n_{vv_1}(k\tau) = 0$$
$$n_{v_1 v_1}(k\tau) = 0$$
$$n_{vu}(k\tau) = \lim_{t \to k\tau, t < k\tau} [n_{vv_1}(t) + n_{v_1 u}(t)]$$
$$n_{v_1 u}(k\tau) = 0$$
$$n_{uu}(k\tau) = \lim_{t \to k\tau, t < k\tau} [n_{vv}(t) + n_{vu}(t) + n_{uu}(t)]$$

which simply reflects the fact that the master sequence has shifted, so that the "$v_1$" chromosomes become the "$v$" chromosomes, the "$v$" chromosomes become "$u$" chromosomes, and the "$u$" chromosomes remain "$u$" chromosomes.

We find that the approximate solutions of the asexual and sexual evolutionary dynamics defined by Eqs. (1) and (2) agree well with stochastic simulations of the exact model (see Figures 1 and 2). Therefore, the approximate model may be used to analyze the asexual and sexual replication pathways being considered in this paper.

After a sufficient number of iterations, the ratios between the various population numbers reach a periodic solution. The criterion for determining whether the asexual population succeeds in adapting on this dynamic landscape is then given by the criterion,

(4) $$\frac{n_{vv}((k+1)\tau)+n_{vu}((k+1)\tau)}{n_{vv}(k\tau)+n_{vu}(k\tau)} > e^{\kappa_{uu}\tau}$$

since the population will only adapt if the organisms with viable chromosomes grow faster than the unviable organisms between the peak shifts[16]. The criterion for adaptability for the sexual population is identical, provided that γρ is sufficiently large that the size of the haploid population is negligible (i.e. if $\gamma\rho\to\infty$ ).

If both the asexual and sexual populations survive on the dynamic landscape, then it is necessary to compare which strategy out-competes the other. This is done by measuring the overall per capita growth of each population in between peak shifts, which is given by,

$$\exp[\frac{1}{\tau}\int_{k\tau}^{(k+1)\tau}\bar{\kappa}(t)dt]$$ , where $\bar{\kappa}(t)$ is the mean fitness of the population, and is simply the first-order growth rate constant of the population as a whole at the given time *t*. The population with the larger per capita growth is the one with the advantageous replication strategy[17].

Figure 3 illustrates regimes where the asexual and sexual populations adapt as a function of both ε and τ. Note that the sexual population always adapts more rapidly than the asexual population (other figures with other parameter values give similar results).

Figure 4 shows plots of the mean fitness as a function of time for both the asexual and sexual populations, for parameter values where both populations adapt according to Figure 3. Note that the sexual population is clearly adapting to the landscape more quickly than the asexual population.

In a static environment, we have previously shown that the asexual mean fitness at steady-state is given by[18],

$$\phi_a = max\{\frac{1}{2}(1+p)^2 - 1, \alpha p\}$$

while the sexual mean fitness at steady-state is given by[18],

$$[\phi_s + 1][\phi_s - \alpha p][\phi_s(1-2\alpha) + \alpha p] - \frac{1}{2}(1+p)^2(1-\alpha)^2 \phi_s^2 = 0$$

where $\phi_a$ and $\phi_s$ are simply the steady-state mean fitnesses of the asexual and sexual populations, respectively, normalized by the fitness of the wild-type. From these expressions, it is possible to show that the sexual population has a higher steady-state mean fitness than the asexual population[18].

The results in this paper and in [18] serve to reconcile two sets of experiments exploring the selective advantage for sex in *S. cerevisiae*[7,11]. In [7], the authors concluded that the purpose of sex is to purge deleterious mutations, while in [11] the authors concluded that sex speeds adaptation to a new environment. Our work suggests that sex serves a dual role, so that both interpretations in [7, 11] are correct.

**Acknowledgments**

The authors would like to thank the United States – Israel Binational Science Foundation (Start-Up Grant) for supporting this research.

P. Gorodetsky derived the ordinary differential equations and numerically implemented them, developed the stochastic codes, and generated the figures. E. Tannenbaum defined the research problem, outlined the approach, guided it to completion, and wrote the paper.

The authors declare that they have no competing financial interests.

Correspondence and requests for materials should be addressed to E.T. (e-mail: emanuelt@bgu.ac.il).

**Figure 1:** Adaptation of the asexual population on a dynamic landscape, simulated using both the approximate system of differential equations and the stochastic simulations. The plot shows the mean fitness of the population as a function of time. Note that the time between peak shifts is sufficiently large that the population is able to reach the steady-state mean fitness ("static landscape" line), which is the same as predicted by theory [18]. Parameter values are $L = 20$, $S = 4$, $\varepsilon = 0.01$, $\tau = 5$, $\kappa_{vv} = 10$, $\kappa_{vu} = 5$, $\kappa_{uu} = 0.1$, and the total population size is 30,000.

**Figure 2:** Adaptation of the sexual population on a dynamic landscape, simulated using both the approximate system of differential equations and the stochastic simulations. The plot shows the mean fitness of the population as a function of time. Note that the time between peak shifts is sufficiently large that the population is able to reach the steady-state mean fitness ("static landscape" line), which is the same as predicted by theory [18]. Parameter values are $L = 20$, $S = 4$, $\varepsilon = 0.01$, $\tau = 5$, $\kappa_{vv} = 10$, $\kappa_{vu} = 5$, $\kappa_{uu} = 0.1$, and the total population size is 30,000.

**Figure 3:** Illustration of the regions of adaptability for both the asexual and sexual populations. The blue continuous line defines the boundary separating the regions where the sexual population adapts, colored black and white, and where the sexual population does not adapt, colored grey. The red line defines an analogous boundary for the asexual population, except that the asexual population only adapts in the white colored region. In the black colored region, the asexual population does not adapt, while the sexual population does. Furthermore, although both the asexual and sexual population adapt in the white region, the sexual population does so more quickly than the asexual population, and so the sexual population out-competes the asexual population in this regime. There is therefore no region where the asexual population out-competes the sexual population. Parameter values are $L = 20$, $S = 4$, $\varepsilon = 0.01$, $\kappa_{vv} = 10$, $\kappa_{vu} = 5$, $\kappa_{uu} = 0.1$, and we assume that $\gamma\rho$ is sufficiently large that the population spends a negligible amount of time in the haploid phase.

**Figure 4:** Comparison of the evolutionary dynamics of the asexual and sexual populations on the dynamic fitness landscape. The plot is of the mean fitness for both populations, generated via stochastic simulations. The parameter regimes are such that both populations adapt to the landscape. However, note that the sexual population adapts more quickly. The parameter values are $L = 20$, $S = 4$, $\varepsilon = 0.01$, $\tau = 2$, $\kappa_{vv} = 10$, $\kappa_{vu} = 5$, $\kappa_{uu} = 0.1$.

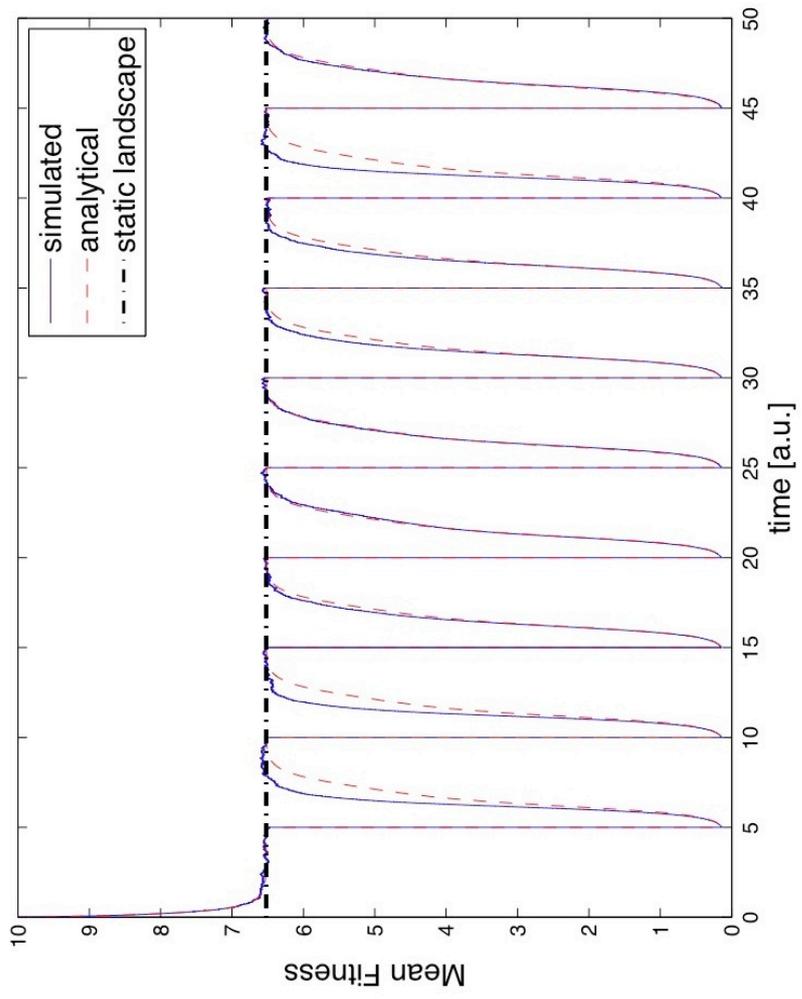

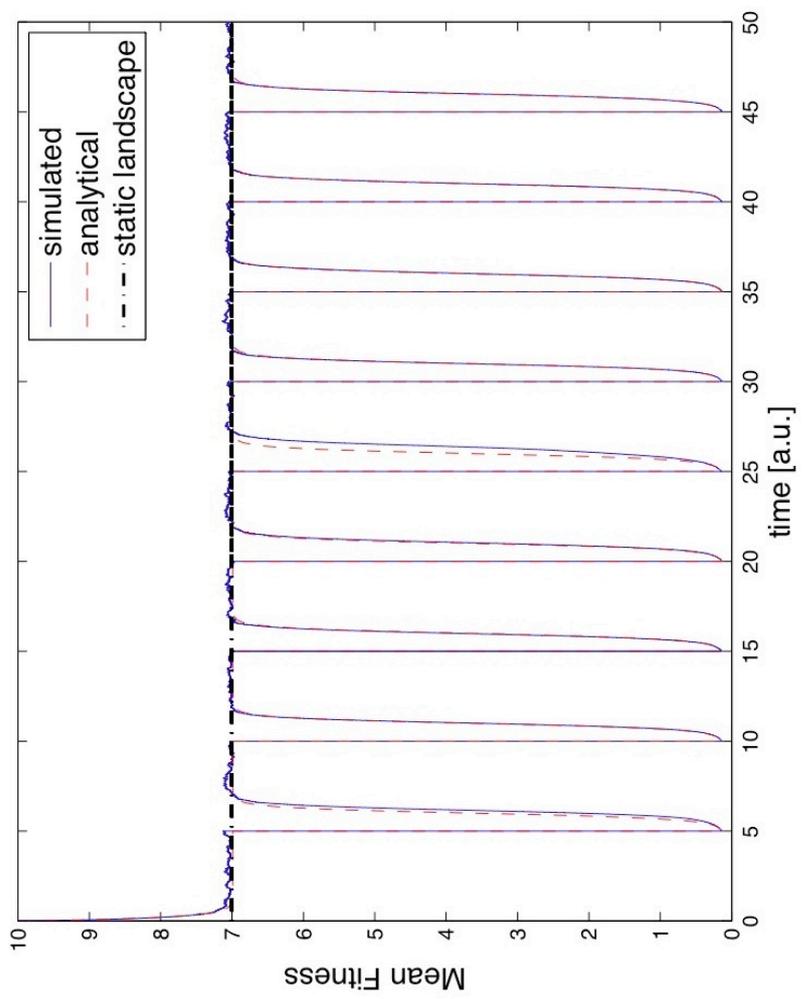

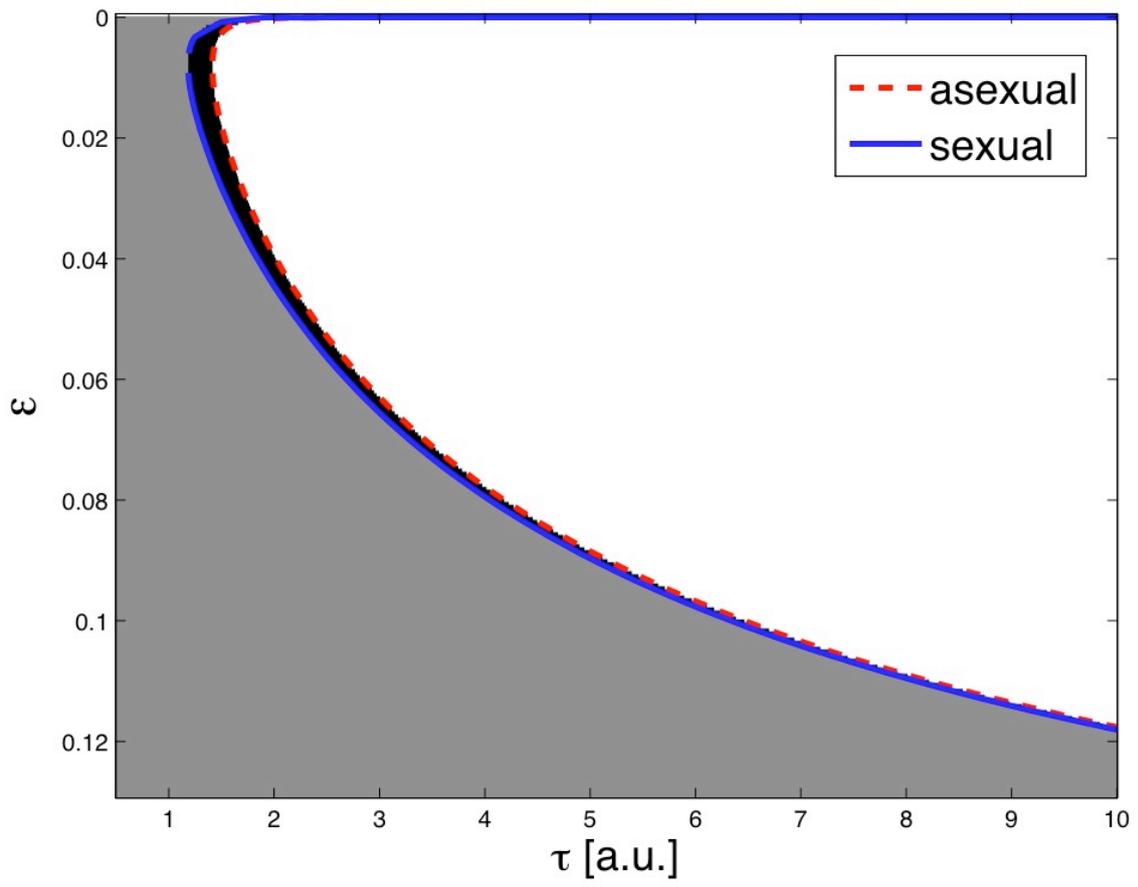

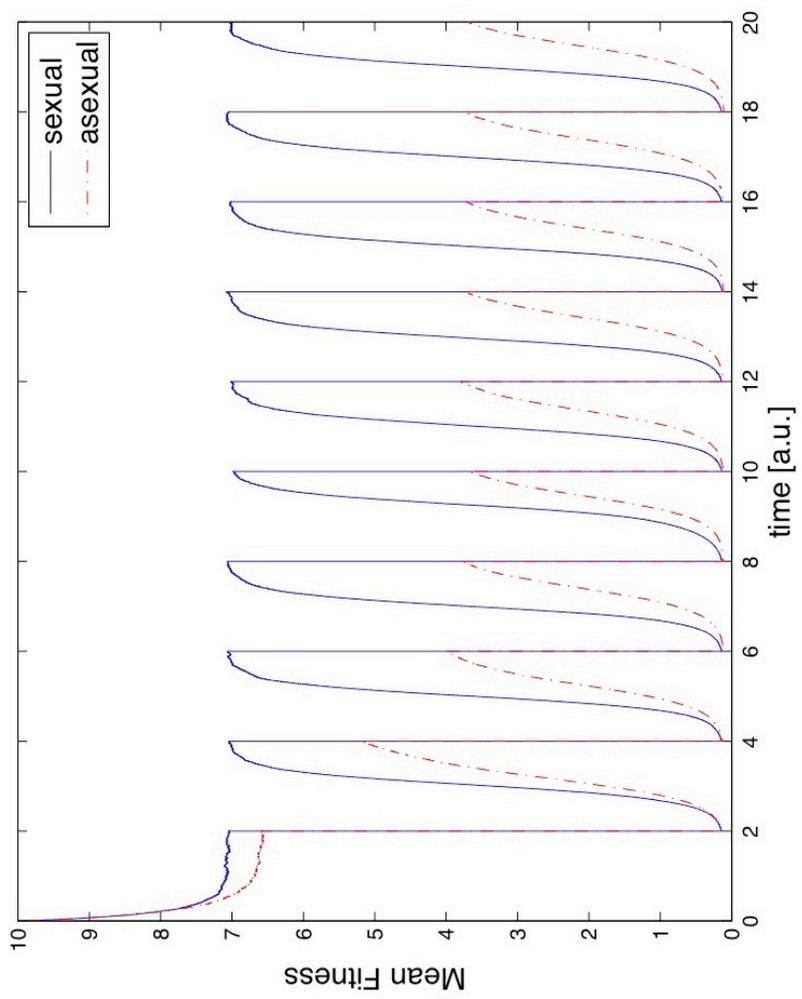